\documentclass[a4paper]{JHEP3}
\usepackage{epsfig}
\title{
\protect\vspace{5mm}
More about spontaneous Lorentz-violation and infrared modification of
gravity }
\author{M.V. Libanov$^{a,b}$ and V.A. Rubakov$^a$\\
$^a$Institute for Nuclear Research of the Russian Academy of
Sciences,\\
60th October Anniversary Prospect 7a, 117312, Moscow, Russia.\\
$^b$ Service de Physique Th\'{e}orique, CP 225,\\
  Universit\'{e} Libre de Bruxelles, B--1050, Brussels, Belgium\\
E-mail: \email{ml@ms2.inr.ac.ru, rubakov@ms2.inr.ac.ru}
}
\preprint{ULB-TH/05-12\\
INR/TH-2005-17}
\abstract{
We consider a model with Lorentz-violating vector field condensates, in
which dispersion laws of all perturbations, including tensor modes,
undergo non-trivial modification in the infrared. The model is free of
ghosts and tachyons at high 3-momenta. At low 3-momenta there are ghosts,
and at even lower 3-momenta there exist tachyons. Still, with appropriate
choice of parameters, the model is phenomenologically acceptable. Beyond a
certain large distance  scale and even larger time scale, the
gravity of a static source changes from that of General Relativity to that
of van~Dam--Veltman--Zakharov limit of the Fierz--Pauli theory. Yet the
late time  cosmological evolution is always determined by the standard
Friedmann equation, modulo small correction to the ``cosmological Planck
mass'', so the modification of gravity cannot by itself explain the
accelerated expansion of the Universe. The latter property
is generic in a wide class of models with condensates. }

\begin{document}
\section{Introduction and summary}
\label{Section/Pg1/1:paper/Introduction and summary}
Recently, there have been several attempts to construct theories in which
gravity gets modified at ultra-large distance and time
scales~\cite{extra,drummond,ghostc,moffat,
Rubakov:2004eb,Dubovsky:2004sg,Dubovsky:2005dw} (see also
Ref.~\cite{Bekenstein:2004ca} and references therein). One of the
motivations is to explain the accelerated expansion of the Universe by a
modification of the Friedmann equation rather than by introducing the
cosmological constant or exotic matter such as quintessence. Most of
the Lorentz-invariant models with unconventional gravity in the infrared
either have ghosts or suffer from the strong coupling problem at
unacceptably low energies~\cite{problems} (see, however,
Ref.~\cite{Nicolis:2004qq}). These problems need not be inherent in
Lorentz-violating
theories~\cite{ghostc,Rubakov:2004eb,gripaios,Dubovsky:2004sg}, so it is
natural to study possible infrared modifications of gravity in the context
of Lorentz-violation. In this paper we consider one model of this sort, to
see what kinds of features may emerge once Lorentz-invariance gets broken.

One way to introduce spontaneous Lorentz-violation is to invoke
condensates of vector (or, more generally, tensor)
fields~\cite{vectora0,moffat2,moffat3,
jacobson,Mattingly,gripaios,Carroll:2004ai,Elling}. Requiring symmetry
under spatial rotations, one can either consider a condensate of the time
component $B_0$ of a vector field $B_\mu
$~\cite{vectora0,moffat2,moffat3}, or introduce internal global symmetry
$G$ and allow for condensates of spatial components which break $G\times
SO(3)_S$ but leave unbroken an $SO(3)$ subgroup, where $SO(3)_S$ is the
original group of spatial rotations. The simplest version of the latter
construction (cf. Ref.~\cite{aia}) involves three vector fields $A_\mu
^a$, $a=1,2,3$, a triplet under $G\equiv SO(3)_G$, whose condensates are
\begin{equation}
A_i^a=\mu \delta _i^a
\label{Eq/Pg1/1:paper}
\end{equation}
where $i=1,2,3$ is the spatial index. Of course, both types of
condensates, $B_0$ and $A_i ^a$, may be present in one and the same model,
and in this paper we consider precisely the latter possibility.

Vector fields condense if their Lagrangian contains a potential term
$V(B_\mu  , A_\mu ^a)$. Hence, the theory of vector fields is definitely
not
gauge-invariant. In this case the kinetic terms in the vector field
Lagrangian are generally not gauge-invariant as well.
Without gauge invariance, there is a danger of ghosts
whose 3-momenta could be unacceptably high. However, this is not
necessarily the case in the presence of vector field
condensates~\cite{gripaios}. In this paper we make use of the mechanism of
Ref.~\cite{gripaios} to avoid ghosts at high momenta.

Violation of Lorentz-invariance by vector field condensates generically
leads to new light modes which are mixtures of gravitational and vector
field excitations. In models considered so far, the dispersion laws
for these modes, as well as for transverse-traceless descendants of gravitons,
is $\omega^2 \propto k^2$, with propagation velocities generally different
for different modes. This possibility has been extensively studied in
literature, and, indeed, it has a number of interesting phenomenological
consequences~\cite{Graesser:2005bg,Elliott:2005va}. We note, however,
that in these models, no  modification of the dispersion laws occur at
ultra-low spatial momenta. This is in contrast to the Higgs mechanism
in gauge theories.
Unlike in gauge theories,
the Lorentz connections are proportional to derivatives of the fundamental
field, the metric. Hence, the terms in the vector field Lagrangian,
quadratic in covariant derivatives, upon condensation lead to
two-derivative terms in the Lagrangian for the metric
perturbations; symbolically
\[
DA\cdot DA\to \mu ^2(\Gamma \cdot \Gamma) \propto \mu ^2 \cdot\partial
h\partial h
\]
where $\Gamma _{\nu \lambda }^\mu $ are Christoffel symbols, and
$h_{\mu \nu }$ are metric perturbations about Minkowski background. On the
other hand, because of general covariance, the potential $V(B, A)$ does
not generate a mass for the graviton, although it may generate mixing
between the graviton and vector field perturbations about their
condensates. Hence, if the vector Lagrangian contains terms with zero and
two derivatives only, the dispersion relation for the graviton
(possibly with admixture of vector fields) is, at low 3-momenta $k$,
\begin{equation}
\omega ^2=\mathrm{ v}^2 \cdot k^2
\label{Eq/Pg2/1(3*):paper}
\end{equation}
where
\[
\mathrm{ v}=1+O\left(\frac{\mu ^2}{M_{Pl}^2} \right)
\]
Unlike in gauge theories with the Higgs mechanism, the disperison law
(\ref{Eq/Pg2/1(3*):paper}) is valid at arbitrarily low spatial momenta.

In this paper we explore a possibility that, besides the potential term
$V(B,A)$ and kinetic two-derivative terms, the vector field Lagrangian
contains terms with one derivative, e.g.,
\begin{equation}
\lambda \varepsilon _{abc}\cdot D_\mu A_\nu ^a A^{b\mu }A^{c\nu }
\label{Eq/Pg2/2(4*):paper}
\end{equation}
where $\lambda \ll 1$. We assume that except for $\lambda $, the vector
field Lagrangian does not contain very small dimensionless parameters, and
that $\mu \ll M_{Pl}$. Upon vector field
condensation~(\ref{Eq/Pg1/1:paper}), the term (\ref{Eq/Pg2/2(4*):paper})
becomes an addition to the quadratic Lagrangian for perturbations of the
vector fields and metric, which is linear in derivatives and is
characterized by the scale $\lambda \mu $. Even without gravity,
introducing the term (\ref{Eq/Pg2/2(4*):paper}) brings in a new feature:
there reappear ghosts, but only at low 3-momenta, $p\lesssim \lambda \mu
$. Unlike in Lorentz-invariant theories, these ghosts --- negative energy
excitations --- are not unacceptable provided that $\lambda \mu $ is low
enough~\cite{Cline,Dubovsky:2004sg} (see also Ref.~\cite{Trodden});
they may even be useful in the cosmological
context~\cite{holdom}. In a certain window of 3-momenta around $p\sim
\lambda \mu $, there may also appear tachyons, which are by far more
dangerous. In the model we consider,  tachyons are absent at the expense
of fine-tuning of the parameters.

Another effect of the term (\ref{Eq/Pg2/2(4*):paper}) in the theory
without gravity is that at $k\ll \lambda \mu $, one of the modes (a
3-scalar) has the dispersion relation
\begin{equation}
\omega ^2 \propto k^4,\ \ \  \ k\ll\lambda \mu
\label{Eq/Pg3/1(5*):paper}
\end{equation}
i.e. it behaves like an excitation of ghost condensate~\cite{ghostc}.

With gravity turned on, two other, even lower scales appear. One of them
is
\begin{equation}
\Lambda _{(1)}=\lambda \mu \cdot \frac{\mu }{M_{Pl}}
\label{Eq/Pg3/2(3*):PAPER}
\end{equation}
This scale determines the \textit{ distance} at which the gravitational
field of a static source gets modified: at $r\ll \Lambda
_{(1)}^{-1}$ this field is the same as in  General Relativity, while
at $r\gg \Lambda _{(1)}^{-1}$ it has the (scalar-tensor) form of the
van~Dam--Veltman--Zakharov limit of the Fierz--Pauli theory,
\begin{equation}
r\ll \Lambda _{(1)}^{-1}:\mbox{ GR} \ \ \longrightarrow\ \  \ r\gg
\Lambda _{(1)}^{-1}:\mbox{ FP}
\label{Eq/Pg3/1(6*):paper}
\end{equation}
Also, at this scale the dispersion relation for the
``ghost-condensate-like'' mode changes from (\ref{Eq/Pg3/1(5*):paper})
back to $\omega^2 \propto k^2$,
\[
k\gg \Lambda _{(1)}:\ \ \omega ^2\propto k^4\ \ \longrightarrow\ \  \
k\ll \Lambda _{(1)}:\ \ \omega ^2\propto k^2
\]
In fact, it is this mode that adds to the gravitational interaction
between the static sources at $r\gg \Lambda _{(1)}^{-1}$.

The lowest scale that appears in our model is
\[
\Lambda _{(2)}=\lambda \mu \frac{\mu ^2}{M_{Pl}^2}
\]
At this scale the behavior of 3-tensor and 3-vector modes changes
considerably. While at $k\gg \Lambda _{(2)}$ the dispersion laws in these
sectors have the form (\ref{Eq/Pg2/1(3*):paper}), at low 3-momenta $(k\ll
\Lambda _{(2)})$ there exist modes with the dispersion law
\[
\omega ^2\propto \pm k
\]
In particular, there are tachyons in the spectrum. They are not
unacceptable, provided that $\Lambda _{(2)}$ is small enough, say,
$\Lambda _{(2)}\lesssim H_0$, where $H_0$ is the present value of the
Hubble parameter. Furthermore, at the very first sight, tachyons are a
welcome feature from the cosmological viewpoint: gravity becomes unstable
 at very large distances and times.

Roughly the same scale determines the \textit{ time} at which
gravitational field of a static source gets modified from GR to FP.
The separation between this time scale and the distance scale,
Eq.~(\ref{Eq/Pg3/2(3*):PAPER}), is basically the same phenomenon as the
one found in the ghost condensate model~\cite{ghostc}. In fact, the time
scale relevant to the modification of Newton's law may be smaller than
$\Lambda _{(2)}^{-1}$. We will keep track of one of the parameters of our
model, call it $\alpha ^B$, and see that the time scale is actually
\[
\frac{1}{\sqrt{\alpha ^B}\Lambda _{(2)}}
\]
Thus,  gravity of static sources may get modified well before the
space becomes unstable\footnote{In this paper we will not discuss
gravity of moving sources, which is expected to have much in common
with that in the ghost condensate model~\cite{star}.}.

We will see, however, that in spite of all these peculiarities, the
Friedmann equation governing the evolution of spatially flat,
homogeneous and isotropic Universe remains precisely the same as in GR,
provided that the vector fields have settled down to their condensate
values (there are light modes that can in principle serve as unusual
particles filling the Universe; we will not consider this possibility in
the present paper). The only difference with respect to GR is that the
``cosmological Newton's constant'' is not exactly equal to the
gravitational constant entering Newton's law at $r\ll \Lambda _{(2)}^{-1}$
--- the phenomenon found in Refs.~\cite{Mattingly,Carroll:2004ai}.
Thus, condensation of vector fields cannot by itself explain the
accelerated expansion of the Universe\footnote{Of course there is always
a (trivial) possibility of non-zero vacuum energy in a state with
condensates of vector fields, which would be nothing but a contribution to
the cosmological constant, see, e.g., Ref.~\cite{moffat3}.}. Furthermore,
we will argue (see also Ref.~\cite{Elling}) that in four-dimensional local
theories allowing for derivative expansion, the form of the gravitational
side of the Friedmann equation for spatially flat Universe, $H^2$, is
fixed by symmetries, so it is unlikely that the accelerated expansion
would occur in four-dimensional theories solely due to the condensation of
any tensor fields. In this regard, brane world scenarios with infinite
extra dimensions appear more promising.

We are not going to discuss phenomenology of our model in this paper.
We only note that for sufficiently small $\mu$ and very small $\lambda \mu$,
this model should be
phenomenologically
acceptable, and that the constraints obtained, e.g., in
Refs.~\cite{Carroll:2004ai,Graesser:2005bg,Elliott:2005va,Cline}
should apply, with appropriate modifications, to our model as well.
The model
displays a number of potentially interesting features --- ghosts at low
3-momenta, tachyons at even lower 3-momenta, modification of the
gravitational fields of sources at large times and not so large distances.
It remains to be understood whether these features may be
phenomenologically useful.

\section{The model}
\label{Section/Pg4/1:paper/The model}

We consider a generally covariant theory with vector fields $B_\mu $ and
$A_\mu ^a$. The latter form a triplet under global symmetry group
$SO(3)_G$. We take the action in the form
\begin{equation}
S=\int \limits_{}^{}\! d^4x\sqrt{-g}\left(\mathcal{ L}_{EH}+\mathcal{
L}_V^{(2)}+\mathcal{ L}_V^{(1)}-V \right)
\label{Eq/Pg4/1:paper}
\end{equation}
where [signature $(+,-,-,-,)$]
\[
\mathcal{ L}_{EH}=-M_{Pl}^2R
\]
is the Einstein-Hilbert term, and the terms $\mathcal{ L}_V^{(2)}$,
$\mathcal{ L}_V^{(1)}$ and $V$ are of the second, first and zeroth order
in derivatives of the vector fields. We will not write down all terms
compatible with the symmetries (we have not found any terms other than
ones displayed below, which would add new features to our model). The term
quadratic in derivatives is basically the same as in Ref.~\cite{gripaios},
except  that we have more fields,
\begin{eqnarray}
\mathcal{ L} _V^{(2)}&=&
-\frac{1}{4g_A^2}F_{\mu \nu }^aF^{a\mu \nu } -\frac{\alpha
^A}{2}(D_\mu A^{a\mu })^2+\frac{\beta _{1}^A}{2M^2}F_{\mu \nu }^aF^{a\mu
\sigma }B_\sigma B^\nu
+\frac{\beta _2^A}{2M^2}D_\mu A^{a\mu }D_\nu
A^{a\sigma }B^\nu B_\sigma
\nonumber\\
&&
 +\frac{\beta _3^A}{2M^2}D_\mu A^{a\nu }D_\sigma
A_\nu ^aB^\mu B^\sigma +\frac{\beta _4^A}{2M^4}D_\mu A^{a\nu }D_\sigma
A^{a\rho }B^\mu B^\sigma B_\nu B_\rho
\nonumber\\
&&
-\frac{1}{4g_B^2}B_{\mu \nu }B^{\mu \nu } -\frac{\alpha
^B}{2}(D_\mu B^{\mu })^2
+\frac{\beta _{1}^B}{2M^2}B_{\mu \nu }B^{\mu
\sigma }B_\sigma B^\nu
+\frac{\beta _2^B}{2M^2}D_\mu B^{\mu }D_\nu
B^{\sigma }B^\nu B_\sigma
\nonumber\\
&&
 +\frac{\beta _3^B}{2M^2}D_\mu B^{\nu }D_\sigma
B_\nu B^\mu B^\sigma +\frac{\beta _4^B}{2M^4}D_\mu B^{\nu }D_\sigma
B^{\rho }B^\mu B^\sigma B_\nu B_\rho
\label{Eqn/Pg5/1:paper}
\end{eqnarray}
where $F_{\mu \nu }^a=\partial _\mu A_\nu ^a-\partial _\nu A_\mu ^a$,
$B_{\mu \nu }=\partial _\mu B_\nu -\partial _\nu B_\mu $. We assume that
the dimensionless constants $g^{A,B},\alpha ^{A,B}$ and $\beta _i^{A,B}$
are all roughly of order 1, and generally do not consider them as small or
large parameters. To simplify formulas below, we will often skip
prefactors depending on these constants. We will use the sign ``$\simeq$''
instead of equality sign in formulas with omitted prefactors. We will keep
track of the constant $\alpha ^B$, since at some point we will consider
what happens if it is larger than others.

The potential term is chosen to be
\begin{equation}
V=\lambda _1  (A_\mu ^aA^{a\mu }+\gamma M^2)^2+\lambda _2(B_\mu B^\mu
-M^2)^2+\lambda _3A_\mu ^aA^{a\nu }B_\nu B^\mu +\lambda _4A_\mu ^aA_\nu
^aA^{b\mu }A^{b\nu }
\label{Eq/Pg6/1(5*):PAPER}
\end{equation}
where again the constants $\lambda _i$ and $\gamma $ are assumed to be
of order 1. Thus, the most important
parameter entering $\mathcal{ L}_V^{(2)}$ and $V$ is the energy scale $M$;
in what follows we assume that $M\ll M_{Pl}$.

Finally, the one-derivative term is
\begin{equation}
\mathcal{ L}_V^{(1)}=
\lambda  D_\mu A_\nu^aA^{b\mu }A^{c\nu }\varepsilon ^{abc}
\label{Eq/Pg5/1(II2*):paper}
\end{equation}
Here
\[
\lambda \ll 1
\]
and in what follows $\lambda $ is considered as small parameter. Note that
this term breaks the symmetry $A_\mu ^a\to -A_\mu ^a$, so its smallness is
technically natural. One can introduce two other one-derivative terms into the
vector Lagrangian, $D_\mu A^a_\nu A^{a \mu} B^\nu$ and
$D_\mu A^a_\nu A^{a \nu} B^\mu$. We have found that at the quadratic level,
the latter terms vanish
for light perturbations,
so we will not consider them in what follows.

The potential term in the Lagrangian gives rise to spontaneous
Lorentz-violation. Indeed, the static configuration
 \begin{eqnarray}
 &&A^{a}_i=\mu \delta ^{a}_i\;,\ \ \ \mu ^2=\frac{\gamma  \lambda _1
 M^2}{3\lambda _1+\lambda _4}\simeq M^2\nonumber
\\
 &&B_0 =M\nonumber
\\
 &&B_i=A_0^a=0\nonumber\\
 &&g_{\mu \nu }=\eta _{\mu \nu }
 \label{Eq/Pg3/1:br}
 \end{eqnarray}
 minimizes the potential and solves the field equations. This background
 is invariant under $SO(3)_S$ spatial rotations complemented with
 $SO(3)_G$ transformations of the global group. Thus, Lorentz-invariance
 is spontaneously violated, while $SO(3)$-invariance stays intact.

 In the next section we will study perturbations of the vector fields
 about the background (\ref{Eq/Pg3/1:br}) in the absence of gravity. A
 general property of these perturbations is that there are neither ghosts
 nor tachyons at sufficiently high 3-momenta, $k\gg \lambda \mu $: with an
 appropriate choice of the parameters $\alpha ^{A,B}, \beta _i^{A,B}$ and
 $g^{A,B}$, the two-derivative terms (\ref{Eqn/Pg5/1:paper}) have
 correct signs~\cite{gripaios}. This does not require fine-tuning; rather,
 a sufficient condition is that the following inequalities are satisfied,
\begin{eqnarray}
 -\alpha ^N+\beta _2^N+\beta _3^N+\beta _4^N&>&0\nonumber\\
 \beta _1^N-\frac{1}{g_N^2}&>&0\nonumber\\
 \frac{1}{g_N^2}-\beta _1^N-\beta _3^N&>&0\nonumber\\
 \frac{1}{g_N^2}&>&0\nonumber\\
 \alpha ^N&>&0
 \label{Eqn/Pg6/1:paper}
 \end{eqnarray}
 separately for $N=A$ and $N=B$. It is straightforward to see that these
 inequalities indeed have a wide class of solutions.

 \section{Spectrum with gravity switched off}
 \label{Section/Pg6/1:paper/Spectrum with gravity switched off}
 \subsection{Preliminaries}
 \label{Subsec/Pg6/1:paper/Preliminaries}

 In what follows we will be interested in the spectrum of perturbations
 about the background (\ref{Eq/Pg3/1:br}). We consider our model as an
 effective field theory valid below the scale $\mu $ (recall that
 $M\simeq \mu $), and study light modes only. We begin with the theory of
 vector fields, with metric perturbations switched off, and write
\begin{eqnarray}
A^{a\mu }&=&-\mu \delta ^{\mu a}+a^{a\mu }\nonumber\\
B^{\mu }&=&M\delta ^{0\mu }+b^\mu
\label{Eqn/Pg3/1:br}
\end{eqnarray}
 In what follows the summation over indices $i$ and $a$ runs with the
 Euclidean three-\-di\-men\-si\-onal metric.

 It is straightforward to see that the symmetric part of $a_i^a$ (obeying
 $a_i^a=a_a^i$), $b_0$ and the combination
\begin{equation}
 Ma_0^i-\mu b_i
 \label{Eq/Pg7/1(III*):paper}
 \end{equation}
 obtain masses of order $\mu $. In the low energy theory they are set
 equal to zero. Thus, light modes can be parametrized by the fields
\begin{eqnarray}
 \alpha _{ij}&=&\frac{a_i^j-a_j^i}{2\mu }\nonumber\\
 \sigma _i&=&\frac{2}{\mu }a_0^i=\frac{2}{M}b_i
\label{Eqn/Pg7/1:PAPER}
\end{eqnarray}
In our conventions, these fields are dimensionless. To study the
 properties of these fields, it is convenient to decompose them into
 transverse vectors and scalars with respect to the unbroken $SO(3)$,
 \begin{equation}
 \alpha _{ij}\equiv -\varepsilon
 _{ijk}d_k-\varepsilon _{ijk}\partial _kd
 \label{Eq/Pg7/2:paper}
\end{equation}
 \[
 \sigma _i\equiv v_i+\partial _i v
 \]
 where $\partial _id_i=\partial _iv_i=0$. We are now able to study vector
 $(v_i,d_i)$ and scalar $(v ,d)$ sectors separately.

 \subsection{Vector modes}
 \label{Subsec/Pg7/1:paper/Vector modes}

 Retaining the terms proportional to $d_i$ and $v_i$ in $(\mathcal{
 L}_V^{(1)}+\mathcal{ L}_V^{(2)})$, we obtain the quadratic Lagrangian for
 vector modes,
 \begin{eqnarray}
 \frac{1}{\mu ^2}\mathcal{ L}^{\mbox{vec}}&=&\frac{\kappa _1}{2}(\partial
 _0d_k)^2-\frac{\kappa _2}{2}(\partial _id_k)^2+\frac{\kappa
 _3}{2}(\partial _0v_k)^2-\frac{\kappa _4}{2}(\partial _iv_k)^2 -\kappa
 _5\varepsilon _{ijk}\partial _iv_j\cdot\partial _0d_k \nonumber \\
 &&+\frac{\lambda \mu }{4}\varepsilon _{ijk}\partial _iv_j\cdot
 v_k+\lambda \mu \varepsilon _{ijk}\partial _id_j\cdot d_k+\lambda \mu
 \partial _0d_i\cdot v_i
 \label{Eqn/Pg8/1:paper}
 \end{eqnarray}
 where $\kappa _n$ are combinations of the parameters entering the
 two-derivative Lagrangian (\ref{Eqn/Pg5/1:paper}).

 At high 3-momenta $k\gg \lambda \mu $, the two-derivative terms dominate,
 and the dispersion relations have the general form
\begin{equation}
 \omega ^2=\mathrm{ v}^2\cdot k^2
 \label{Eq/Pg8/1:paper}
\end{equation}
 where $\mathrm{ v}^2$ are different for different modes, and different
(maybe, substantially different) from 1. All of them are positive if the
 inequalities (\ref{Eqn/Pg6/1:paper}) are satisfied; $\kappa _1,\ldots
 ,\kappa _4$ are positive as well, so that there are neither ghosts nor
 tachyons at $k\gg \lambda \mu $.

 To study the case of general 3-momenta, one performs Fourier
 transformation and introduces two circular polarizations labeled by
 $\epsilon =\pm 1$, so that the $\epsilon $-modes of, say, the field
 $v_i$ obey $\varepsilon _{ijk}k_j v_k^{(\epsilon )}=i\epsilon k
 v_i^{(\epsilon )}$. Modes with different polarizations decouple, and the
 equation determining the dispersion relation is
\[
 \left[\kappa _1\omega ^2-(\kappa _2k^2-\epsilon k\lambda \mu )   \right]
 \left[\kappa _3\omega ^2-(\kappa _4k^2-\epsilon k\lambda \mu )   \right]=
 \omega ^2(\kappa _5\epsilon k+\lambda \mu )^2
 \]
 In general, there is a range of momenta for which one of the two modes
 with $\epsilon >0$ becomes tachyonic; in this range the combination
\begin{equation}
 (\kappa _2k-\lambda \mu )(\kappa _4k-\lambda \mu )
 \label{Eq/Pg8/1(III4*):paper}
 \end{equation}
 is negative. To avoid tachyons, we fine tune the parameters,
 \begin{equation}
 \kappa _2=\kappa _4
 \label{Eq/Pg8/2:paper}
 \end{equation}
 thus ensuring that the combination (\ref{Eq/Pg8/1(III4*):paper}) is
 semi-positive-definite\footnote{The requirement of the absence of tachyons
 imposes also a certain inequality between the parameters, which is easy
 to satisfy.}. With this fine-tunig, all four frequencies are real; their
 behavior as function of $k$ is shown in Fig.~\ref{Figure1}. At small $k$
the dispersion relations are
\EPSFIGURE[t]{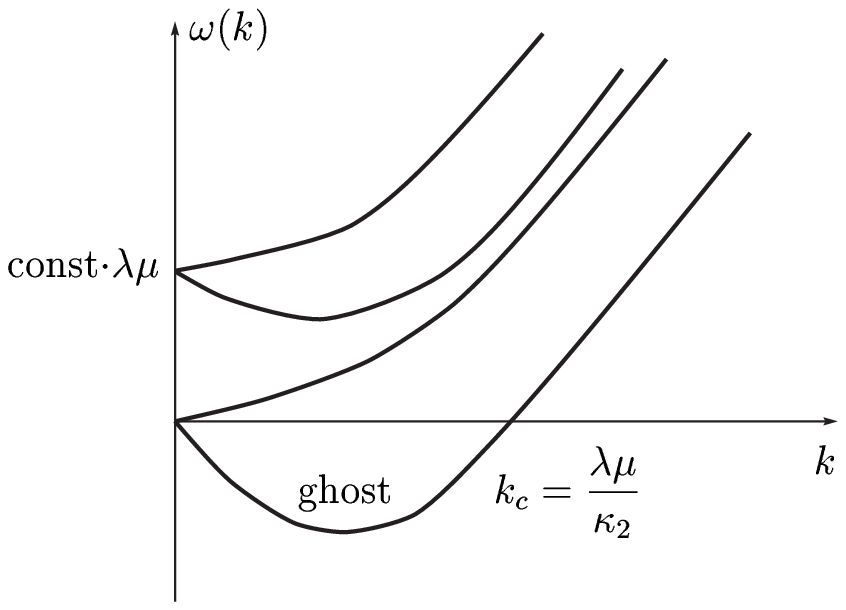,width=7.9cm}{
\label{Figure1}Dispersion relations $\omega (k)$ for
vector modes. Not shown are four other frequencies which are mirror
symmetric against $\omega =0$, i.e., $\omega \to -\omega $.
\vspace{0cm}
}
\begin{eqnarray}
\omega ^2(k)&\simeq& \left[
\begin{array}{l}
(\lambda \mu) ^2\\
k^2
\end{array}
\right.\ \ \ \  k\ll\lambda \mu
\label{Eqn/Pg9/1:paper}
\end{eqnarray}
and at
\[
k=k_c\equiv \frac{\lambda \mu }{\kappa _2}\simeq \lambda \mu
\]
one of the modes crosses zero and becomes a gapless ghost. The latter fact
is obvious from quantum-mechanical viewpoint; it can be also checked by an
explicit calculation of the classical energy of a wave: one finds that the
energy is negative at $k<k_c$. As discussed in
Refs.~\cite{Cline,Dubovsky:2004sg},
this is phenomenologically acceptable provided that $k_c$ is small enough.

\subsection{Scalar sector}
\label{Subsec/Pg9/1:paper/Scalar sector}

The analysis of the scalar sector is even simpler. We write for light modes
\[
a_0^i=\frac{\mu }{2} \partial _iv, \ \
b_0^i=\frac{M}{2} \partial _iv
\]
where $v$ is one of the two light fields, the second being the field $d$.
These two fields decouple  in the two-derivative Lagrangian $\mathcal{
L}_V^{(2)}$, while they mix through the one-derivative term $\mathcal{
L}_V^{(1)}$. The quadratic Lagrangian is then
\begin{equation}
\frac{1}{\mu ^2}\mathcal{ L}^{\mbox{s}}=\nu _1(\partial _0\partial
_id)^2-\nu _2(\partial _i^2d)^2+\nu _3(\partial _0\partial _iv)^2-\nu
_4(\partial _i^2v)^2-2\lambda \mu \partial _0d\cdot \partial _i^2v
\label{Eq/Pg9/1:paper}
\end{equation}
where $\nu _i$ are again combinations of the parameters entering
$\mathcal{ L}_V^{(2)}$. At $k\gg\lambda \mu $ the dispersion relations
again have the form (\ref{Eq/Pg8/1:paper}), while at $k\ll \lambda \mu $
the two scalar modes have quite different dispersion
relations\footnote{Keeping track of $\alpha ^B$, the relevant momentum
scale here is $\lambda \mu /\sqrt{\alpha ^B}$. This is not important
for our discussion.}: one is
\[
\omega ^2\simeq(\lambda \mu )^2,\ \ \ \ k\ll\lambda \mu
\]
and another is
\begin{equation}
\omega ^2=\frac{\nu _2\nu _4}{(\lambda \mu )^2} k^4,\ \ \ \ k\ll\lambda
\mu
\label{Eq/Pg10/1(*8):PAPER}
\end{equation}
In what follows it will be instructive to keep track of the parameter
$\alpha ^B$, assuming that it is larger than the other parameters in
$\mathcal{ L}_V^{(2)}$. It enters $\nu _2$ only, so the dispersion
relation (\ref{Eq/Pg10/1(*8):PAPER}) is, in fact,
\begin{equation}
\omega ^2\simeq\frac{\alpha ^B}{(\lambda \mu )^2}k^4,\ \ \  \ k\ll \lambda
\mu
\label{Eq/Pg9/1(a*):paper}
\end{equation}
As before, the sign ``$\simeq$'' means that we omit a prefactor of order 1.

Notice that the dispersion relation (\ref{Eq/Pg9/1(a*):paper}) is similar
to the dispersion relation in the ghost condensate model, but, unlike in
the latter, it holds at low 3-momenta only. Note also that there are
no ghosts or tachyons in the scalar sector even at $k\ll \lambda \mu $.

\section{Spectrum with gravity}
\label{Section/Pg10/1:paper/Spectrum with gravity}
\subsection{Light fields}
\label{Subsec/Pg10/1:paper/Light fields}

Once gravity is turned on, in addition to perturbations of the vector
fields, Eq.~(\ref{Eqn/Pg3/1:br}), one considers also metric perturbations,
$g_{\mu \nu }=\eta _{\mu \nu }+h_{\mu \nu }$.
Under the gauge transformations of the general covariance, the
perturbations transform as follows,
\begin{eqnarray}
\delta a^{a\mu }&=&\mu \partial _a\xi ^\mu \nonumber\\
\delta b^\mu &=&-M\partial _0\xi ^\mu
\nonumber\\
\delta h_{\mu \nu }&=&\partial _\mu \xi _\nu +\partial _\nu \xi _\mu
\nonumber
\end{eqnarray}
There are three local gauge-invariant combinations with no derivatives,
$(\mu h_{ij}-a_i^j-a_j^i)$, $(Mh_{00}+2b_0)$ and $(M\mu h_{0i}-Ma_0^i+\mu
b_i)$. It is these combinations that can enter the quadratic part of the
potential $V$. Indeed, one finds at quadratic order
\begin{equation}
V\!=\!\lambda _1\mu ^2(\mu h_{ii}-2a^i_i)^2\!+\lambda _4\mu ^2(\mu
h_{ij}-a_i^j-a_j^i)^2\!+\lambda _2M^2(Mh_{00}+2b_0)^2+\!\lambda _3
\left(M\mu h_{0i}-Ma^i_0+\mu b_i\right)^2
\label{Eq/Pg3/2(C):br}
\end{equation}
Thus, the light fields can be parametrized by the five combinations, two
given by (\ref{Eqn/Pg7/1:PAPER}) and others being $h_{ij}, h_{0i}$ and
$h_{00}$, while light components of other fields are expressed through
these five (e.g., $a_i^j+a_j^i=\mu h_{ij}$).

The decomposition of $h_{\mu \nu }$ similar to (\ref{Eq/Pg7/2:paper}),
reads
\begin{eqnarray}
h_{ij}&=&\chi _{ij}+\partial _is_j+\partial _js_i+\partial _i\partial
_j\sigma +\delta _{ij}\tau \nonumber\\
h_{0i}&=&u_i+\partial _i\zeta \nonumber\\
h_{00} &= & \varphi\nonumber
\end{eqnarray}
where $\chi _{ij}$ is transverse-traceless, while $s_i$ and $u_i$ are
transverse. The convenient choice of gauge is
\begin{equation}
 s_i=0,\ \ \ \ \zeta =v=0
 \label{Eq/Pg11/1:paper}
 \end{equation}
 By gauge invariance of the original action, the same number of fields
 (one transverse vector and two scalars) must be
 non-propagating\footnote{The fields $s_i,\zeta $ and $v$ transform as
 $\delta s_i=\xi _i^\perp$, $\delta \zeta =\xi _0+\partial _0\xi
 ^\parallel$ and $\delta v=2\xi_0$, where $\xi _i^\perp$ and $\partial
 _i\xi ^\parallel$ are transverse and longitudial parts of the gauge
 function $\xi _i$. Thus, the gauge choice (\ref{Eq/Pg11/1:paper}) is
 almost Hamiltonian, and there are no spurious propagating degrees of
 freedom.}; these are $u_i$, $\varphi $ and $\sigma $. The former two
 enter the action without time derivatives, while $\sigma \simeq
 (1/\partial _i^2)\tau $ due to equations of motion.

 The analysis of the spectrum is tedious but straightforward. Let us
 summarize the results.

 \subsection{Tensors}
 \label{Subsec/Pg11/1:paper/Tensor sector}

 In the tensor sector, the quadratic Lagrangian is
 \[
 \mathcal{ L}^{\mbox{\tiny T}}=\frac{M_0^2}{2}(\partial _0\chi
 _{ij})^2-\frac{M_1^2}{2}(\partial _k\chi _{ij})^2+\frac{\lambda \mu
 ^3}{4}\varepsilon _{abi}\partial _i\chi _{ka}\cdot\chi _{kb}
 \]
 where  $M_0^2,M_1^2=M_{Pl}^2+O(\mu ^2)$, and the last term is due to the
 one-derivative part $\mathcal{ L}_V^{(1)}$ in the vector Lagrangian.
 Thus, in the tensor sector, only the scale $\Lambda _{(2)}=\lambda \mu
 ^3/M_{Pl}^2$ is relevant. Above this scale, the dispersion law is almost
 standard,
 \[
 \omega ^2=\mathrm{ v_{\mbox{\tiny T}}}^2\cdot k^2,\ \ \  \ k\gg \Lambda
_{(2)}
\]
 where $\mathrm{ v_{\mbox{\tiny T}}}=1+O(\mu ^2/M_{Pl}^2)$. Below the scale
$\Lambda _{(2)}$, the dispersion relation is
\[
 \omega ^2\simeq \pm \Lambda _{(2)}\cdot k,\ \ \  \ k\ll \Lambda
 _{(2)}\equiv \frac{\lambda \mu ^3}{M_{Pl}^2}
 \]
 Minus sign here corresponds to tachyon. One of the tensor modes
 becomes unstable at $k\simeq \Lambda _{(2)}$.

 \subsection{Vectors}
 \label{Subsec/Pg11/1:paper/Vector tensor}

 In the vector sector, one eliminates the non-dynamical field $u_i$ by
 making use of its equation of motion. Then a new term appears in the
 quadratic Lagrangian,
\begin{equation}
 \frac{1}{\mu ^2}\triangle \mathcal{ L}^{\mbox{vec}}\simeq -\frac{\lambda
 ^2\mu ^4}{M_{Pl}^2}v_i^2
 \label{Eq/Pg12/1:paper}
 \end{equation}
 The coefficients of already existing terms (\ref{Eqn/Pg8/1:paper})
 receive corrections of order $\mu ^2/M_{Pl}^2$. The latter property is
 not harmless: to avoid tachyons in finite, albeit narrow, momentum range
 around $k\sim \lambda \mu $, one has to modify the fine-tuning relation
 (\ref{Eq/Pg8/2:paper}) by $M_{Pl}$-suppressed corrections.

 The term (\ref{Eq/Pg12/1:paper}) becomes comparable to one-derivative
 terms in (\ref{Eqn/Pg8/1:paper}) at $k\sim \Lambda _{(2)}$. Therefore, the
 spectrum (\ref{Eqn/Pg9/1:paper}), with light ghost, persists down to the
scale $\Lambda _{(2)}$. At lower momenta the dispersion relations of the
gapless modes are
\[
\omega ^2\simeq \mp \Lambda _{(2)}\cdot k,\ \ \ \ k\ll\Lambda _{(2)}
\]
while the other two modes still have $\omega ^2=(\lambda \mu )^2$. The
ghost existing in the vector sector at $k\lesssim \lambda \mu $ becomes
tachyonic at $k\simeq \Lambda _{(2)}$.

\subsection{Scalars}
\label{Subsec/Pg12/1:paper/Scalar sector}

The Lagrangian in the scalar sector, in the gauge (\ref{Eq/Pg11/1:paper}),
is
\begin{eqnarray}
\mathcal{ L}^{\mbox{s}}&=&\nu _1\mu ^2(\partial _0\partial _id)^2-\nu _2\mu
^2(\partial _i\partial _kd)^2+\frac{1}{4}\nu _3\mu ^2(\partial _i\varphi
)^2-\frac{1}{4}\nu _4\mu ^2(\partial _i^2\partial _0\sigma )^2+\lambda \mu
^3\partial _id\cdot\partial _i\varphi \nonumber\\
&-&2\lambda \mu ^3\partial
_id\cdot\partial _i\tau -3\tilde{M}_1^2(\partial _0\tau
)^2+\tilde{M}_2^2(\partial _i\tau )^2-2\tilde{M}_3^2\partial _i\tau\cdot
\partial _i\varphi -2\tilde{M}_4^2\partial _0\tau \cdot\partial _0\partial
_i^2\sigma
\label{Eqn/Pg12/2(f*):paper}
\end{eqnarray}
where $\nu _i$ are the same as in (\ref{Eq/Pg9/1:paper}), and
$\tilde{M}_i^2=M_{Pl}^2+O(\mu ^2)$. The field $\varphi $ is not dynamical
indeed, while $\partial _i^2\sigma $ is algebraically expressed through
$\tau $ via its equation of motion. While the gauge
(\ref{Eq/Pg11/1:paper}) is convenient for calculating the spectrum in the
general case, the limit $M_{Pl}\to \infty $ is somewhat tricky. One way to
proceed is to express $\sigma $ and $\tau $ through $\varphi $ and $d$
using $\sigma $- and $\varphi $-equations, and then plug these $\sigma $
and $\tau $ into $d$- and $\tau  $-equations. The resulting system for
$d$ and $\varphi $ coincides, in the limit $M_{Pl}\to \infty $, with the
field equations obtained from the Lagrangian (\ref{Eq/Pg9/1:paper}), if one
identifies $\varphi $ with $(-2\partial _0v)$.

The spectrum corresponding to the Lagrangian (\ref{Eqn/Pg12/2(f*):paper})
coincides with that discussed in section \ref{Subsec/Pg9/1:paper/Scalar
sector} down to the momentum scale $\Lambda _{(1)}=\lambda \mu ^2/M_{Pl}$.
At this scale the dispersion relation of the soft mode
(\ref{Eq/Pg9/1(a*):paper}) changes to
\begin{equation}
\omega ^2\simeq \frac{\alpha ^B\mu ^2}{M_{Pl}^2}\cdot k^2,\ \ \ \
k\ll\Lambda _{(1)}\equiv \lambda \frac{\mu ^2}{M_{Pl}}
\label{Eq/Pg13/1:paper}
\end{equation}
These waves propagate in the usual way, but their velocity is small.

To end up this section, we note that the parameter $\alpha ^B$ appears in
 the scalar sector only; neither the momentum scales nor dispersion
 relations depend on $\alpha ^B$ in the tensor and vector sectors.

\section{Gravity of static source}
\label{Section/Pg13/1:paper/Gravity of static source}

With the Lagrangian (\ref{Eqn/Pg12/2(f*):paper}), it is straightforward to
couple $h_{00}\equiv \varphi $ to static source $T_{00}$ and calculate the
gravitational field. The result is
\begin{eqnarray}
\varphi &=&\frac{c_1M_{ Pl}^2\triangle-4\lambda ^2\mu ^4}
{c_1M_{ Pl}^2\triangle-3\lambda ^2\mu ^4}\times\frac{1}{2\hat{M}_{
Pl}^2\triangle}T_{00}\nonumber\\
\tau &=&\varphi +\frac{2\lambda ^2\mu ^4}{c_2M_{Pl}^2\triangle-3\lambda
^2\mu ^4}\times\frac{1}{2\hat{M}_{
Pl}^2\triangle}T_{00}
\label{Eqn/Pg13/1:paper}
\end{eqnarray}
where $\triangle\equiv \partial _i^2$, $c_1, c_2= 1+O(\mu ^2/M_{Pl}^2)$ and
$\hat{M}_{Pl}^2=M_{Pl}^2+O(\mu ^2)$. These potentials coincide with the GR
expressions at $r\ll \Lambda _{(1)}^{-1}$ (with the ``Newton's law Plank
mass'' $\hat{M}_{Pl}$), and at large \textit{ distances} tend to the
(scalar-tensor) expressions of the Fierz--Pauli theory in the vDVZ limit,
\[
\begin{array}{l}
\varphi =\displaystyle\frac{2}{3}\frac{1}{\hat{M}_{Pl}^2\triangle
}T_{00}\\ \\
\tau  =\displaystyle\frac{1}{3}\frac{1}{\hat{M}_{Pl}^2
\triangle}T_{00}
\end{array}
 ,\ \ \  \ r\gg \Lambda _{(1)}^{-1}
 \]
The corresponding time scale can be obtained from the dispersion relation
(\ref{Eq/Pg13/1:paper}), which at $k\sim \Lambda _{(1)}$ gives
\begin{equation}
\omega ^{-1}(k\sim \Lambda _{(1)})\simeq \frac{1}{\Lambda _{(2)}\sqrt{\alpha ^B}}
\label{Eq/Pg14/1(h*):paper}
\end{equation}
For $\alpha ^B\sim 1$ this scale coincides with the time scale at which
the tachyonic instability develops in the tensor and vector sectors. If,
on the other hand, one takes $\alpha ^B$ large, the time scale
(\ref{Eq/Pg14/1(h*):paper}) at which GR is violated gets smaller than the
tachyonic time scale $\Lambda _{(2)}^{-1}$. Modification of Newton's law may
occur earlier than the flat space becomes unstable.

\section{Cosmological evolution}
\label{Sec/Pg14/1:paper/Cosmological evolution}

Finally, let us consider our model in the cosmological context, assuming
that the Universe is spatially flat. The metric is
\begin{equation}
ds^2=N^2(t)dt^2-a^2(t)dx_i^2
\label{Eq/Pg14/2(j*):paper}
\end{equation}
The potential (\ref{Eq/Pg6/1(5*):PAPER}) does not have flat directions
respecting the residual $SO(3)$ symmetry, so at late times it is
consistent to set the vector fields to their condensate values (in locally
Minkowski frame), i.e.,
\[
B_0=N(t)\cdot M
\]
\[
A_i^a=a(t)\cdot\mu \delta _i^a
\]
Plugging these expressions back into the action (\ref{Eq/Pg4/1:paper}),
one evaluates the action in terms of the lapse function $N(t)$ and scale
factor $a(t)$. One finds that all terms in the action have one and the
same form, so that the total action is
\begin{equation}
S^{\mbox{\tiny cosm}}={\left(M_{Pl}^{\mbox{\tiny cosm}} \right)^2}\int
\limits_{}^{}\!dt \frac{1}{N}\left(\frac{\dot{a}}{a} \right)^2
\label{Eq/Pg14/3(j*):paper}
\end{equation}
where $(M_{Pl}^{\mbox{\tiny cosm}} )^2=M_{Pl}^2+O(\mu ^2)$. The form of
the action (\ref{Eq/Pg14/3(j*):paper}) coincides with that of GR, so the
cosmological evolution in our model is governed by the standard Friedmann
equation. The only peculiarity is that the ``cosmological Planck mass''
$M_{Pl}^{\mbox{\tiny cosm}} $ in general does not coincide with the
``Newton's law Planck mass'' $\hat{M}_{Pl}$ entering
(\ref{Eqn/Pg13/1:paper}); this property has been observed in
Refs.~\cite{Mattingly,Carroll:2004ai} in a similar context.

This result appears to be of rather general nature (cf. Ref.~\cite{Elling}),
at least in
four-dimensional theories obeying locality in time and allowing for
derivative expansion of the action.
The spatially flat, homogeneous and
isotropic metric is symmetric under time reparametrizations and space
dilations,
\[
t\to t' (t)
\]
\[
 x^i\to \lambda  x^i
\]
With matter fields settled down at their vacuum values (in locally
Minkowski frame), the only terms in the action which are local in time,
consistent with these symmetries and have not more than two time
derivatives are (\ref{Eq/Pg14/3(j*):paper}) and $\int \limits_{}^{} Ndt$,
the latter having the same form as the cosmological constant term. No
matter what condensates are there in the Universe, its evolution proceeds
according to the Friedmann equation (possibly with modified Newton's and
cosmological constants), provided that the condensates are
time-independent (in locally Minkowski frame) and consistent with
homogeneity and isotropy of space.

 This observation is not entirely trivial. In our model, the
 energy-momentum tensor of the vector condensates, calculated by making
 use of the vector Lagrangian~(\ref{Eqn/Pg5/1:paper}),
 (\ref{Eq/Pg6/1(5*):PAPER}), (\ref{Eq/Pg5/1(II2*):paper}), is non-zero.
 The point is that in the spatialy flat, homogeneous and isotropic
 Universe it is necessarily proportional to the Einstein tensor (and in
 general also $g_{\mu \nu }$ and terms with higher time derivatives). This
 results in the appearance of $M_{Pl}^{\mbox{\tiny cosm}}$ instead of
 $M_{Pl}$ in the action~(\ref{Eq/Pg14/3(j*):paper}).

\acknowledgments
 The authors are indebted to S.~Dubovsky,
D.~Gorbunov, M.~Luty, R~Rattazzi, S.~Sibiryakov, P.~Tinyakov, and
I.~Tkachev for helpful comments. We thank Service de Physique
Th\'{e}orique, Universit\'{e} Libre de Bruxelles, where this work has
been completed, for hospitality. This work has been supported in part by
RFBR grant 05-02-17363-a, by the Grants of the President of Russian
Federation NS-2184.2003.2, MK-3507.2004.2, by INTAS grant YSF 04-83-3015,
by the grant of the Russian Science Support Foundation
and  by contract IAP V/27 of IISN, Belgian Science Policy.

\end{document}